\documentclass{aastex}
\usepackage{emulateapj5}
\usepackage{psfig, epsfig}

\newcommand{\gtae}{$\buildrel {\lower3pt\hbox{$>$}} \over 
{\lower2pt\hbox{$\sim$}} $}
\newcommand{\ltae}{$\buildrel {\lower3pt\hbox{$<$}} \over
{\lower2pt\hbox{$\sim$}} $}

\slugcomment{to appear in ApJ, vol. 564, Jan 2002}

\begin{document}

\def\xte1550{XTE~J1550$-$564} 

 
\title{The 1998 Outburst of XTE~J1550$-$564: \\
     A Model Based on Multiwavelength Observations }


\author{K.~Wu\altaffilmark{1,2}, R.~Soria\altaffilmark{1}, 
  D.~Campbell-Wilson\altaffilmark{2}, D.~Hannikainen\altaffilmark{3}, 
  B.~A.~Harmon\altaffilmark{4}, R.~Hunstead\altaffilmark{2}, 
  H.~Johnston\altaffilmark{2}, M.~McCollough\altaffilmark{4},  
  and V.~McIntyre\altaffilmark{2,5}} 
\email{kw@mssl.ucl.ac.uk; rs1@mssl.ucl.ac.uk}
\altaffiltext{1}{MSSL, 
  University College London, Holmbury St.~Mary, Dorking, Surrey, RH5~6NT, UK}  
\altaffiltext{2}{School of Physics, University of Sydney, 
  NSW 2006, Australia} 
\altaffiltext{3}{Department of Physics and Astronomy,
  University of Southampton, Southampton, SO17~1BJ, UK} 
\altaffiltext{4}{SD50, NASA-MSFC, Huntsville, AL 35812, USA} 
\altaffiltext{5}{ATNF, CSIRO, PO Box 76, Epping, NSW 1710, Australia}

\begin{abstract}
The 1998 September outburst of the black-hole 
X-ray binary \xte1550 was monitored 
  at X-ray, optical and radio wavelengths. We divide the outburst sequence 
into five phases and discuss their multiwavelength properties.  
The outburst starts with a hard X-ray spike, 
while the soft X-ray flux rises with a longer timescale. 
We suggest that the onset of the outburst is determined 
by an increased mass transfer rate from the companion star, 
but the outburst morphology is determined by the distribution 
of specific angular momentum in the accreting matter. 
The companion in \xte1550 is likely to be an active magnetic star, 
  with a surface field strong enough 
  to influence the dynamics of mass transfer. 
We suggest that its magnetic field can create a magnetic bag 
  capable of confining gas inside 
  the Roche lobe of the primary. 
The impulsive rise in the hard X-rays is explained by the inflow 
   of material with low angular momentum  
   onto the black hole, on a free-fall timescale, 
   when the magnetic confinement breaks down. At the same time, 
high angular momentum matter, transferred via ordinary Roche-lobe 
overflow, is responsible for the formation of a disk.
 
\end{abstract}




\section{Introduction}

Many soft X-ray transients (SXTs) are binaries containing a low-mass
  star transferring material to a black hole.
SXT outbursts are usually attributed 
  to accretion-disk or mass-transfer instabilities, and 
  show a variety of patterns and properties.  
It has been suggested that the companion star 
may also play an important role, depending on 
its evolutionary status and its response to irradiative heating.    
(See e.g.\ Tanaka \& Lewin 1995 for a review.) 
The diversity of the outburst behavior in SXTs 
  may therefore arise from the different nature of their companion stars.

\xte1550 was discovered on MJD~51063 
  (1998 Sep 7; MJD$=$JD$-$2,400,000.5) 
  by the All-Sky Monitor (ASM) on board 
  {\it RXTE} (Smith 1998) 
  and by the Burst and Transient Source Experiment (BATSE) 
  on board {\it CGRO} 
  (Wilson et al.\ 1998).   
The optical counterpart was identified shortly afterwards
  (Orosz, Bailyn \& Jain 1998), 
  and a radio source was found at the optical position 
  (Campbell-Wilson et al.\ 1998).
From the optical ellipsoidal modulations, 
an orbital period of $1.541\pm0.009$~d was inferred (Jain et al.\ 2001).  
For a detailed analysis of the X-ray spectral behavior of the source 
during the 1998 outburst, based on RXTE observations, 
see Sobczak et al.\ (2000).

\section{X-ray and radio observations}

The BATSE experiment onboard {\it CGRO} (Fishman et al. 1989) was used to
monitor the hard X-ray emission from XTE J1550$-$564.  The BATSE Large Area
Detectors (LADs) can monitor the whole sky almost continuously in the energy 
range of 20 keV -- 2 MeV with a typical daily 3 sigma sensitivity of better
than 100 mCrab.  Detector counting rates with a timing resolution of 2.048
seconds are used for our data analysis.  To produce the XTE J1550$-$564
light curve, single step occultation data were taken using a standard Earth 
occultation analysis technique used for monitoring hard X-ray sources 
(Harmon et al. 1992).  Interference from known bright sources was removed.  
A spectral analysis of the BATSE data indicated that the data were well 
fitted by a power law with a spectral index of $-3.0$.  
The single occultation step data 
were then fitted with a power law with this index to determine daily flux 
measurements in the 20--100 keV band.

The initial detection of XTE J1550$-$564 as a radio source was made on 
MJD 51065 with the Molonglo Observatory Synthesis Telescope (MOST) at 
843 MHz (Campbell-Wilson et al. 1998).  The evolution of the radio 
source was monitored with MOST over the following 27 days, using 
partial synthesis observations with integration times ranging from 1 
to 9 hours; calibration sources were PKS B1421$-$490 and B1934$-$638. 
Flux density measurements were complicated by sidelobes from the 
nearby radio-bright supernova remnant G326.3-1.8 (Whiteoak \& Green 
1996), necessitating an image differencing approach using a matched 
reference image of the field obtained earlier in 1998 when XTE 
J1550$-$564 was quiescent.  The flux density estimates are given in 
Table 1, with the errors being the quadrature combination of a 3 mJy 
rms noise contribution and a 3\% calibration uncertainty.  Other radio 
data obtained at higher frequencies with the Australia Telescope 
Compact Array, together with VLBI images from the Australian Long 
Baseline Array, will be presented elsewhere (Hannikainen et al. 2002, in 
preparation); preliminary results are given in Hannikainen et al. (2001).

\section{The 1998 Outburst Sequence}  
  
Figure 1 shows the UV/optical/IR, hard/soft X-ray and radio lightcurves 
  of \xte1550 between MJD~51050 and 51155.  The data in the top panel 
of Fig. 1 are from Jain et al. (1999), and Sanchez-Fernandez et al. (1999).   
We divide the 1998 outburst into five phases,
named after the corresponding stages  
  in the {\it RXTE}/ASM X-ray light curve 
  (bottom panel of Fig.~1): 
  (1) fast rising, (2) slow rising,  
  (3) flare, (4) post-flare plateau, and 
  (5) X-ray decline.     

{\it Phase 1: fast rise (MJD~51063--51064.5)}   
The onset of the outburst is characterised by 
  an impulsive rise in the hard X-rays.  
The BATSE (20$-$100~keV) flux 
  reached about 0.3~ph~cm$^{-2}$s$^{-1}$ within one day, 
  and peaked at about 0.4~ph~cm$^{-2}$s$^{-1}$ the next day. 
From {\it RXTE}/PCA and {\it RXTE}/HEXTE data (Wilson \& Done 2001), 
the 20$-$200~keV flux peaks at 
$\approx 1.7 \times 10^{37}~{\rm erg}~{\rm s}^{-1}$, for a distance of 
2.5 kpc (Sanchez-Fernandez et al.\ 1999).
Assuming a ``canonical'' efficiency $\eta \sim 0.1$, 
  the total accreted mass required to account for the initial 
  hard X-ray burst is \ltae $10^{23}$~g. In fact, 
such high values of $\eta$ are more appropriate for standard 
disk accretion. Hot advective flows in the inner region, 
where the hard X-rays are likely to be emitted, are 
radiatively inefficient, especially if the inflowing  
matter carries a large radial velocity.
For $\eta \sim 10^{-3}$--$10^{-4}$
(Quataert \& Narayan 1999), the required mass would be 
$10^{25}$--$10^{26}$~g.

The soft X-rays showed an exponential-like rise: 
  the {\it RXTE}/ASM (2$-$12~keV) count rate increased 
  from 2 to 40~ct~s$^{-1}$, 
  with an e-folding timescale \ltae~$0.5$~d.  
The optical brightness reached $(V,I) \approx (16.7,14.6)$ 
  (Jain et al.\ 1999),  
  with the possibility that peak brightness had occurred 
  before the optical counterpart was identified.   
The optical colours were redder than in the other phases, 
  but the X-ray spectrum was harder (Wilson \& Done 2001).  
Archival data from the MOST 
  at 843~MHz showed no obvious detection of radio emission 
  before the outburst.

{\it Phase 2: slow rise (MJD~51064.5--51074)}    
This phase can be divided into two stages, 
  (A) MJD~51064.5--51068.5 and (B) 51068.5--51074, 
  based on the rising timescales of the soft X-rays.  
The {\it RXTE}/ASM count rate increased from 40 to 120~ct~s$^{-1}$
  with an e-folding timescale $\approx 3$~d in stage A, 
slowing down markedly through stage B with an e-folding timescale
\gtae~12~d.  In contrast, the hard X-ray flux dropped rapidly:
  it had fallen to around half of the Phase 1 peak value 
  by the end of stage A (MJD~51068.5),    
  and remained essentially constant throughout stage B. 
Table 2 lists the hard and soft X-ray luminosity 
over Phases 1 and 2, inferred from the {\it RXTE}/PCA data 
(Wilson \& Done 2001).

Radio emission became detectable, 
  with MOST flux densities at a level of 10--30~mJy.  
Optical emission remained steady, 
  with $(U,B,V,I) \approx (18,18,17,15)$  
  (Sanchez-Fernandez et al.\ 1999; Jain et al.\ 1999).  

     
{\it Phase 3: flare (MJD~51074--51080)}   
There was a strong flare in all 
wavebands.  
Within one day the {\it RXTE}/ASM count rate rose rapidly 
  from 200 to 500~ct~s$^{-1}$.  
After peaking on MJD~51076, 
  it dropped back to the 200-ct~s$^{-1}$ level in one day,
  and then to 100 ct~s$^{-1}$ in another two days.  
The BATSE flux increased almost simultaneously 
  with the soft X-rays,    
  reaching a value of $\sim$~0.5~ph~cm$^{-2}$~s$^{-1}$ 
  within a day.  
It then dropped back to the 0.2-ph~cm$^{-2}$~s$^{-1}$ level 
  in the next day.  
Both the {\it RXTE}/ASM and BATSE lightcurves appear to be symmetric 
  with respect to the flare peak in this phase,  
  but differences in the rising and falling times 
  are revealed in higher time-resolution BATSE data.     

The radio flare peaked at MJD~51077.8$\pm0.1$, $\sim 1.8$~d 
  after the X-ray peak, 
  with the flux density reaching 380~mJy at 843~MHz.  
The decay was slower than in the X-ray bands, 
  taking four days to decline to the 100-mJy level, 
  which is still about 5 times the pre-flare flux density.  
The optical flare showed a delay of approximately one day 
  with respect to the X-ray flare, peaking on MJD~51077.05; 
  the brightening was less than 0.8~mag 
  in each of the $B$, $V$ and $I$ bands. 
  The optical emission reddened during the flare 
  (Jain et al.\ 1999).

{\it Phase 4: post-flare plateau (MJD~51080--51107)}   
The {\it RXTE}/ASM count rate showed some scatter around  
  a mean level of $\sim$~120~ct~s$^{-1}$,
while the BATSE flux remained    
steady  at the 0.2-ph~cm$^{-2}$~s$^{-1}$ level.  
The 843~MHz flux density continued to decline monotonically, 
  reaching $\sim$~14~mJy at MJD~51092, 
  at which point the monitoring was stopped.  
The $B$, $V$ and $I$ brightness of the source faded steadily
  by $\approx 0.04$~mag~d$^{-1}$ (Jain et al.\ 1999).  

{\it Phase 5: X-ray decline (MJD~51107--51150)}   
Both the {\it RXTE}/ASM  
  and the BATSE fluxes declined,      
  with e-folding timescales $\approx$ 11 and 9~d respectively. 
The optical brightness of the source continued to fade, 
  with $B$ falling below 19.5~mag on MJD~51113 
  (Jain et al.\ 1999).

\section{Magnetic Nature of the Companion Star}   

The 1.54-day orbital period 
  together with the faintness of the quiescent optical counterpart
  suggest that \xte1550 is a low-mass system.  
The companion star should be close to filling its Roche lobe, 
  so that the mass transfer can be rapid enough  
  to fuel the outbursts and sustain the subsequent soft state.  
The mean density of the companion star  
  is therefore $\approx$ 0.077~g~cm$^{-3}$, 
  similar to that of early B main-sequence stars or G/K subgiants.   
The quiescent brightness and the observed colour differences 
  favor a late-type subgiant.

There are two consequences of having a late-type subgiant companion 
  in the system.  
Firstly, if we adopt a 1.5-day orbital period,
  and assume that a 10-M$_\odot$ black hole is 
  powering isotropic emission of X-rays 
  at 0.1 of the Eddington luminosity,     
  the fraction of X-ray luminosity 
  ($\sim 10^{36}~{\rm erg}~{\rm s}^{-1}$) 
  intercepted by the companion star  
  will be much greater than the intrinsic luminosity 
  of the G/K subgiant companion 
  ($\sim 10^{33}$--$10^{35}~{\rm erg}~{\rm s}^{-1}$).   
Irradiation by soft X-rays will cause 
  surface evaporation and blanketing of the energy    
  diffusing from the stellar interior.  
Irradiation by hard X-rays will result in the deposition
  of energy deep in the stellar envelope 
  (e.g.\ Vilhu, Ergma \& Fedorova 1994).  
A cool subgiant companion is therefore susceptible 
  to irradiative heating and consequent instabilities. 
  
Secondly, while the tidally-deformed stellar envelope 
  can be locked in synchronous rotation with the orbit, 
  the degenerate stellar core 
  may not attain perfect rotational synchronism.
The differential rotation leads to dynamo action and magnetic activity,  
  similar to the situation in the RS CVn systems, 
  which are magnetic close binaries containing a G/K subgiant.
For instance, in the well-studied RS CVn system HR~1099, with a
2.8-day orbital period, there is strong evidence (Donati et al.\ 1992,
Vogt et al.\ 1999) that the K1~IV component harbours a multi-kilogauss
global dipolar field.


Given that fast rotating isolated G/K subgiants 
  (Schrijver \& Pols 1993) and G/K subgiants in RS CVn systems 
  are highly magnetically active, 
  G/K subgiants in X-ray binaries can also be magnetic.  
A surface field strength of 1~kG, 
  corresponding to a magnetic stress of
  $4\times 10^4$~erg~cm$^{-3}$, 
  is strong enough to influence 
  the mass- and energy-transport process in the stellar atmosphere.  
By postulating a magnetic G/K subgiant companion  
for \xte1550
  we are implying a new scenario   
  in which the outburst properties 
  are determined not only by accretion-disk instabilities, 
  mass-transfer instabilities and
  irradiative heating, but also by stellar magnetism. 

\section{The phenomenology of the outburst}  
Three remarkable features of the 1998 outburst of \xte1550 are: 
  (1) an impulsive rise in the hard X-rays, 
      while the soft X-ray flux remained low 
      at the onset of the outburst; 
  (2) a subsequent giant flare that occurred almost simultaneously 
      in both hard and soft X-rays 
      about twelve days after the initial rise of the hard X-rays; 
  (3) an exponential decay in the
      soft X-ray brightness at the end of Phase 5, 
      with the same timescale as those seen in
      the 1999 and 2000 outbursts.
     
\subsection{Impulsive inflow and hard X-ray burst}   

It is generally accepted that hard X-ray emission  
is not thermal emission from an accretion disk (whose maximum 
temperature is $\sim 1$ keV), but is 
produced by inverse Compton scattering of softer photons 
off highly energetic ($E \sim 100$ keV) electrons near the BH. 
Theoretical studies have shown that accretion of matter 
with low angular momentum
  tends to produce hard X-rays 
  (Igumenshchev, Illarionov \& Abramowicz 1999; 
  Beloborodov \& Illarionov 2000; 
  Chakrabarti \& Titarchuk 1995).  
The physical justification is that free-falling 
electrons in a quasi-spherical inflow 
can acquire kinetic energies \gtae 100 keV as they approach 
the BH horizon. The kinetic energy of the infalling electrons 
powers the Comptonisation process, either directly (bulk-motion 
Comptonisation) or after it has been converted into thermal energy 
(thermal Comptonisation).


Impulsive hard X-ray bursts can occur 
  if the injection of material is also impulsive 
  and with a narrow distribution of angular momenta.   
We therefore argue that an impulsive injection of matter with 
  low angular momentum 
  was responsible for the initial hard X-ray spike 
  in the 1998 outburst, 
  and that this initial condition  
  could have been a consequence of a magnetic companion star. 
 
\subsection{Magnetic bags around G/K stars}
Extended regions of cool, optically-thin, magnetically confined material 
  around a K star are a common feature of RS CVn systems. 
The typical scale-height of these prominences above the photosphere 
  is about twice the stellar radius, 
  implying a volume of $\sim 10^{34}$--$10^{35}$ cm$^3$ 
  for the confining region.  
Prominences evolve over timescales of days,  
   and contain at any given time a mass of $\sim 10^{20}$ g 
   (Hall \& Ramsey 1994). 

We argue that the mass of cold, magnetically-confined gas 
   can in fact be even higher in \xte1550, 
   when the magnetic companion star is close to 
   filling its Roche lobe. 
The inner Lagrangian (L$_1$) point is a saddle point  
   where the gravitational force is negligible,
   and matter can be more easily lifted up 
   from the photosphere of the companion star. 
Moreover, the gas is prevented from moving in any direction 
  other than along the axis between the stars; at   
  the same time, the magnetic field of the companion star 
   restricts the flow along that axis. 
The gas mass builds up (over a timescale of a few days) 
   until eventually the gas pressure overcomes the magnetic barrier
(see Figure 2 for a cartoon picture of our model).

For magnetic confinement near L$_1$, 
  the magnetic tension/stress must be greater than the gas pressure, 
  implying a critical density for the confined matter:   
\begin{equation}  
  \rho_{\rm c} \approx  {\mu {m_{_{\rm H}}B^2}\over {8\pi kT}}    
        \sim 2 \times 10^{-8} {\rm g~cm}^{-3} 
        \biggl({B \over {1 ~{\rm kG}}} \biggr)^2  
        \biggl({T \over {10^4~{\rm K}}}\biggr)^{-1} \ . 
\end{equation} 
For the sizes of the confining regions in RS CVn systems, 
and if we assume uniform density, 
  this corresponds to a mass upper limit of $\sim 10^{26}$~g. 
(If the gas accumulates near 
the loops' apices, the total mass may be 
less than this value when the confinement breaks down.) 
This is consistent with the accreted mass that is required 
to explain the initial hard X-ray burst (\S 3).
By analogy with RS CVn systems, 
  the confined gas in \xte1550 
  may be located well beyond the L$_1$ point, 
  deep into the Roche lobe of the primary, 
  but corotating with the binary. 
The mass loss rate of the secondary would then 
determine the time necessary to fill the magnetic bags.

At the start of an outburst, as the magnetic companion 
star is close to filling its Roche lobe, 
   part of the photospheric layers is lifted 
   and confined in a magnetic bag beyond the L$_1$ point 
   (low angular momentum component of the accretion flow), 
   while another component 
   starts to be transferred through the Lagrangian point      
  (high angular momentum component). 
It is the low-angular momentum component 
   that we believe is responsible for the peculiar  
   hard X-ray spike seen in Phase 1, while the high-angular 
   momentum component leads to the formation of the disk.

\subsection{Quasi free-fall accretion}
As the magnetic prominence evolves, 
  the material confined beyond the L$_1$ point tends 
  to collect in the loop apex, 
  until its density increases and it eventually bursts 
  the magnetic dam, breaking free of the field. 
In a frame corotating with the orbit, 
the average initial angular momentum of the accreting matter 
  with respect to the black hole is low, i.e., it 
is effectively in radial free-fall.  
In contrast, the angular momentum of the matter accreted 
  via conventional Roche-lobe overflow is large,
  because the matter passing through the L$_1$ point 
  is transonic, 
  and the accretion stream has a non-zero pitch angle 
  (see Lubow \& Shu 1975).
  
The low angular momentum component of the flow will deviate 
  from near-radial 
  only when approaching the black hole.    
When the infalling material 
  encounters its centrifugal barrier in the vicinity of the hole, 
  an accretion shock may be formed, 
  converting the kinetic energy to radiation.   
The photons would then be upscattered 
  by the hot electrons, causing an outburst of hard X-rays.  
In an alternative scenario, the photons may be upscattered 
as they interact with the bulk motion of the infalling electrons. 
When the low-angular-momentum accreting material is depleted, 
  the quasi-spherical inflow subsides,    
  and the hard X-ray luminosity declines.    
At the same time, an accretion disk is formed at large radii 
  by the matter transferred via Roche-lobe overflow.
The steady building and expansion of the disk inwards via diffusion processes 
  results in a gradual increase of the optical brightness 
  and then of the soft X-ray luminosity, evident in Phases 1 and 2.

The {\em{hard}} X-rays can penetrate deep
  into the star, and deposit energy into the convective envelope, 
  thus disturbing the thermal equilibrium of the star.
The star must readjust its structure in response to the sudden heating,
  and it will expand until a new equilibrium is established.
At the same time, {\em soft} X-ray irradiation   
  causes an instantaneous heating of the upper stellar atmosphere, 
  and hence a rapid increase in the atmospheric scale height.  
The timescale for thermal readjustment of
  the stellar envelope (up to a few weeks) is longer than 
  the dynamic and thermal timescales of atmospheric activity  
  induced by soft X-ray irradiation (Vilhu et al.\ 1994). 
Although both hard 
  and soft X-ray irradiation will eventually 
  cause an increase in the rate of mass 
  transfer, there will be a delayed response 
  from the companion star 
  to the initial strong hard X-ray heating.
The subsequent onset of a soft state 
  is probably due to the subgiant regaining contact 
  with its critical Roche surface,  
  as a result of the delayed expansion caused 
  by the hard X-ray heating. 

\subsection{Trigger of the giant flare}     

During Phases 1 and 2 
  we have accretion of free-falling matter with low angular momentum 
  at small radii, 
  and accretion of matter with high angular momentum, 
  via Roche-lobe overflow, at large radii. 
The former creates 
  a hot, pressure-supported, sub-Keplerian ``disk'' 
  or quasi-spherical region at small radii, 
  growing outwards; the latter creates a Keplerian ring 
  slowly diffusing inwards and outwards, 
  forming a standard Shakura-Sunyaev disk.
A discontinuity arises 
when the two accretion flows meet, 
  and the inner boundary of the Keplerian disk  
  is subject to a strong shear.  
We suggest that this leads to a sudden increase in mass transfer 
  as the outer, denser, colder gas suddenly loses 
  part of its angular momentum at this shear discontinuity, 
  thereby triggering the giant soft and hard X-ray flare around MJD~51076 
  (Phase 3).


   
The initial hard X-ray burst in Phase 1 
  and the giant flare in Phase 3 have different soft X-ray properties.  
Moreover, the latter was accompanied by strong radio activity. 
Because of insufficient shear,   
  the quasi-spherical accretion flow in Phase 1 
  is unlikely to amplify the seed magnetic field 
  carried by the infalling matter significantly.  
Only after an accretion disk is formed will 
  dynamo action become efficient and amplify the field.  
One possibility is that
  when the magnetic disk joins the sub-Keplerian inner region, 
  and mass accretion suddenly increases (giant flare), 
  the magnetic field is dragged rapidly 
  towards the black-hole event horizon by the accreting matter. 
The field would be compressed and might reconnect,
  causing the expulsion of disk material 
  and hence the formation of a radio flare/jet.  

The small increase in optical brightness
during the giant X-ray flare is surprising.
However, the heating of the companion star and of the accretion disk 
  could have saturated due to strong X-ray irradiation during 
  the previous phases, and the accretion disk may already have been 
  extended to its outer tidal limit.
 
\subsection{Drainage of the accretion disk}  
Around MJD~51110, both the hard and soft X-ray luminosity   
  began to decline exponentially,
  with the soft X-rays lagging the hard.  
The e-folding decay timescale is $11.0\pm0.1$~d 
  for {\it RXTE}/ASM and slightly shorter for BATSE.   
The corresponding decay timescales for the later outbursts 
in 1998/1999 and 2000 
  are almost identical: 
  $11.3\pm0.1$ and $11.1\pm0.2$~d respectively (Figure~3).

The decline in the X-ray luminosity 
  is therefore evidence of the drainage of the accretion disk.    
The timescale on which a disk is emptied is  
\begin{equation} 
  t \approx \frac{R_{\rm d}^2}{\nu} \sim 
     \frac{1}{\nu} \left(\frac{0.6}{1+q}\right)^2
       \left[ \left(\frac{P}{2\pi}\right)^2
      GM_1(1+q) \right]^{2/3}  
\end{equation}   
  (Pringle 1981), 
  where $G$ is the gravitational constant, 
  $P$ the orbital period, 
  $M_1$ the black-hole mass, 
  $q~(\equiv M_2/M_1)$ the mass ratio, 
  $R_{\rm d}$ the disk radius, and 
  $\nu$ the effective viscosity. 
A decay timescale of 11 days implies 
   $\nu \sim 2.3 \times 10^{17}~{\rm cm}^2~{\rm s}^{-1}$      
   if $q \approx 0.1$, $M_1 = 10$~M$_\odot$. 
The effective viscosity is often expressed in terms of a constant 
parameter $\alpha$ (Shakura \& Sunyaev 1973): 
\begin{equation}  
\nu \approx \alpha H c_{\rm s} \approx \alpha c_{\rm s}^2/\Omega,
\end{equation} 
where the sound speed $c_{\rm s} \approx (kT/m_{\rm p})^{1/2}$, and 
the Keplerian angular velocity 
$\Omega \approx 1.2 \times 10^{-3}$ ($M/10\,$M$_\odot$)$^{1/2}$ 
($R/10^{11}$ cm)$^{-3/2}$ s$^{-1}$. 
For temperatures $10^4$ \ltae \ $T$ \ltae \ $10^5$ K, 
typical of the outer regions 
of the disk ($R \approx R_{\rm d}$), the observed timescale 
corresponds to an effective 
$\alpha \sim 10$ at large radii.

\section{A model for the onset of the outbursts}

\subsection{A subclass of BHC outbursts} 

A remarkable characteristic of the outburst behaviour of \xte1550  
  is the presence of a sharp hard X-ray spike during the initial rise.    
This spectral behaviour is different to that 
  observed in systems 
  such as GRO~J1655$-$40, GS~1354$-$64 or GX~339$-$4. 
The 1996 outburst of GRO~J1655$-$40 started with a very soft state 
  (Hynes et al. 1997), and  
  the hard X-rays turned on about a month after the start of the outburst.  
In GS 1354$-$64, the soft and hard X-ray fluxes are well correlated  
  during their rise and decline (Brocksopp et al. 2001). 
In GX 339$-$4, the soft and hard fluxes tend to be anticorrelated 
  (Corbel et al. 2000).  
No initial hard X-ray spike is detected in any of those three systems. 
  
In fact, \xte1550 is not the only system showing a hard X-ray spike 
  at the onset of an outburst.   
XTE J1859$+$226 
  (Wood et al. 1999; McCollough \& Wilson 1999; Brocksopp et al. 2002) 
  and XTE J2012$+$381 (Vasiliev et al.\ 2000) are two other good examples.  
In these three systems, 
  the hard X-ray flux reached a maximum within the first day after detection, 
  and then declined after about 4 days.
The soft X-ray flux increased more steadily and at a slower pace, 
  for $\sim 10$ days. 
The similarity in the X-ray spectral properties 
  of \xte1550, XTE J1859$+$226 and XTE J2012$+$381 
  can be seen from their hardness-ratios plots (Figure 4; 
  data from the public {\it RXTE}/ASM archive).     
These systems probably belong to a subclass of BH transients,  
  sharing a similar physical mechanism 
  that gives rise to a strong hard X-ray spike  
  at the onset of an outburst.  
A more detailed analysis of the hard X-ray properties 
  of this type of systems will be presented elsewhere  
  (Soria et al 2001, in preparation).

\subsection{Formation of the hard X-ray spike}

We propose that the difference 
  between the outburst behaviour of systems like \xte1550  
  and of the other BH transients is due to the difference in 
  the relative angular momentum content of their accreting matter.  
Matter accreted via a stellar wind has a lower specific angular 
momentum than matter accreted via Roche-lobe overflow, 
  and it has been suggested that 
  the wind-fed systems and systems with Roche-lobe overflow  
  have different preferential spectral states  
  (Beloborodov \& Illarionov 2001; Wu et al.\ 2001).  
There is also evidence that  
  the accretion flow in BH X-ray binaries 
  consists of separate components, 
  each with a different distribution of angular momentum. 
Observations of 
  1E~1740$-$294, GRS~1758$-$258 and GX~339$-$4 
  (Smith, Heindl \& Swank 2001) indicate that 
  the delay of the soft component relative to the hard component 
  (approximately equal to the difference between the viscous 
  and the free-fall timescale) is more evident   
  for systems with Roche-lobe overflow, 
  i.e.\ those where a large accretion disk is present,    
  and the high angular momentum component dominates. 
Shorter viscous delays were observed   
  for wind-fed systems 
  (e.g.,\ Cyg X-1 in the hard state; Smith et al.\ 2001)  
  where the low-angular momentum component dominates.   
 
We attribute the presence of the hard X-ray spike in \xte1550 
  to an inflow of accreting matter with low angular momentum, 
  lasting over the first few days.      
The 3--200 keV X-ray spectrum of \xte1550 
  at the peak of the initial hard phase (Wilson \& Done 2001) 
  is in fact very similar 
  to the spectrum of Cyg X-1 in its hard state,  
  consistent with the intepretation 
  that the initial accretion flow in \xte1550 was dominated 
  by the low angular-momentum component.

\subsection{Outburst morphology: role of the specific angular momentum}

The hard and soft states of BHC outbursts are usually explained by models 
  invoking an optically thick, geometrically thin accretion disk 
  which is initially truncated in the inner region.   
The optical/IR flux comes mostly from the outer disk region.
The soft X-ray flux rises on a viscous timescale 
  as the disk grows inwards at the beginning of the outburst, 
  and is therefore delayed with respect to the optical flux.
At small radii, the accretion flow is hot and quasi-spherical.
The hard X-rays are due to inverse-Compton scattering 
  of soft photons from the disk or the companion star 
  by highly energetic electrons in the hot quasi-spherical flow 
  near the BH. 
When the quasi-spherical flow subsides 
  and is replaced by the thin accretion disk extending down towards 
  the innermost stable circular orbit of the BH, 
  the system is observed in a high/soft state.

In our model, the outburst behaviour is determined by the mass-transfer rate 
  (which determines the energetics of the burst) 
  and the angular momentum distribution of the accreting matter 
  (which determines the initial spectral properties and the X-ray rise time).  
The inner, quasi-spherical flow is fed 
  by the accretion of material with low angular momentum, 
  while the disk is formed by accretion of 
  high angular momentum matter  
  transferred via Roche-lobe overflow.  
The initial hard X-ray emission is caused 
  by a sudden injection of accreting matter with low angular momentum.  
When the injection ends and the 
  supply of low-angular momentum accreting matter is depleted,  
  the hard X-ray flux declines.  
In the case of \xte1550, we attribute the decline of the low angular 
momentum accretion flow
  to the emptying of the magnetic reservoir.  
After the magnetic bags are drained, 
  accretion proceeds only via Roche-lobe overflow. 

Our model is in contrast to scenarios 
  based on advection-dominated accretion flows 
  (ADAF models; see Narayan \& Yi 1995; Esin et al.\ 1997), 
  in which spectral softening  
  is attributed to the collapse of the hot, quasi-spherical flow 
  into an optically thick disk 
  when the mass accretion rate increases. 
An application of the ADAF model to the 1998 outburst of \xte1550 
has been presented by Wilson \& Done (2001).
The conventional ADAF scenarios do not explain why in some systems 
  (such as \xte1550) the hot inner region produces 
  a strong hard X-ray spike before collapsing, 
  while in other systems (such as GRO~J1655$-$40) 
  no such initial spike is observed. 
Moreover, the observed duration of the initial hard emission 
  seems to be much longer than 
  the timescale for the collapse of the hot flow 
  in the conventional ADAF model (Wilson \& Done 2001).

\section{Summary}  

We describe the morphology of the 1998 outburst of \xte1550, dividing it 
into five phases.  
The outburst starts with a hard X-ray flare (peaking after $\sim 1$ d); 
the same behaviour is observed in at least two other BHCs.
We propose that the companion star in \xte1550 
  is magnetically active, and 
its magnetic field creates a magnetic bag 
  capable of confining $\sim 10^{26}$ g of gas inside 
  the Roche lobe of the primary, 
  corotating with the binary.
The impulsive rise in the hard X-rays is explained
   by the inflow 
   of material with low angular momentum  
   onto the black hole, on a free-fall timescale, 
   when the magnetic confinement breaks down.  
At the same time, matter with higher angular momentum,  
  transferred via Roche-lobe overflow,  
  begins to form a Keplerian disk, responsible for 
  the soft X-rays and optical emission. 
We suggest that the onset of the outburst is determined 
by an increased mass transfer rate from the companion star, 
but the outburst morphology is determined by the distribution 
of specific angular momentum in the accreting matter.
When the inner boundary of the outer, denser 
  Keplerian disk comes into contact with the inner sub-Keplerian hot region, 
  the disk matter is subject to a strong shear and loses angular momentum. 
This disturbs the stability of the accretion disk, causes a sudden increase 
  in the mass accretion rate, and is responsible 
   for the observed giant X-ray flare and radio ejections.
An identical timescale of 11 days is found 
   for the decay of the soft X-ray flux 
   in the three outbursts observed in this system,    
   implying an effective viscosity  
   $\nu \sim 2.3 \times 10^{17}~{\rm cm}^2~{\rm s}^{-1}$ 
   for the accretion disk.

\acknowledgments

We thank Celia Sanchez-Fernandez and Raj Jain for allowing us to use   
their optical data, and Colin Wilson for providing helpful information 
on his X-ray spectral fitting. We also thank the anonymous referee 
for providing useful suggestions on how to improve our paper.
KW acknowledges support from the ARC Australian Research Fellowship 
   and the PPARC-MSSL Visting Fellowship, and 
   thanks Phil Charles 
   for funding his visits to the University of Southampton.  
RS acknowledges support from the University of Sydney during his visit there.
DCH acknowledges the support of a UK PPARC postdoctoral research grant 
  to the University of Southampton, 
  and financial support from the Academy of Finland, 
  and thanks the Astrophysics Department, Sydney University 
  for hospitality during her visits. 
MOST is operated by Sydney University and supported by 
  grants from the Australian Research Council.





\clearpage

\begin{deluxetable}{cc}
\tabletypesize{\scriptsize}
\tablecaption{MOST flux densities for XTE J1550$-$564 during the 1998 
outburst}
\tablewidth{0pt}
\tablehead{
\colhead{Time at mid-observation (MJD) } & \colhead{Flux density at 843 MHz 
(mJy)}
}
\startdata
51065.175  & $12 \pm 3$\\
51066.156  & $ 16 \pm  3$\\
51071.203  & $ 27 \pm  3$\\
51073.193  & $ 18 \pm  3$\\
51076.390  & $168 \pm  6$\\
51077.298  & $327 \pm  10$\\
51078.259  & $375 \pm  12$\\
51079.318  & $221 \pm  7$\\
51080.302  & $155 \pm  6$\\
51081.210  & $120 \pm  5$\\
51087.360  & $ 20 \pm  3$\\
51092.177  & $ 14 \pm  3$
\\
 \enddata

\end{deluxetable}

\clearpage

\begin{deluxetable}{cccc}
\tabletypesize{\scriptsize}
\tablecaption{X-ray luminosity of \xte1550 during Phases 1-2}
\tablewidth{0pt}
\tablehead{
\colhead{Obs.\ date (MJD) } & \colhead{$L_{\rm 3-20~keV}$ 
($10^{37}~{\rm erg}~{\rm s}^{-1}$)}   
& \colhead{$L_{\rm 20-200~keV}$ ($10^{37}~{\rm erg}~{\rm s}^{-1}$)}   &
\colhead{$L_{\rm bol}$ ($10^{37}~{\rm erg}~{\rm s}^{-1}$)}
}
\startdata
51063 & 0.67 & 1.67 & 3.41 \\
51064 & 0.84 & 1.57 & 3.15\\
51065 & 1.49 & 1.59 & 4.41\\
51067 & 2.24 & 1.17 & 5.38\\
51072 & 2.96 & 0.83 & 8.13
\\
 \enddata


\tablecomments{{\it RXTE}/PCA plus {\it RXTE}/HEXTE data, 
from Wilson \& Done (2001). We assumed a distance of 2.5 kpc 
(Sanchez-Fernandez et al. 1999).}

\end{deluxetable}

\clearpage


\begin{figure}
\vspace{-1.2cm}

\hspace{0.645cm} \epsfig{figure=f1a_col.eps,width=9.08cm,angle=270}\\

\vspace{-0.8cm}
\hspace{0.18cm} \epsfig{figure=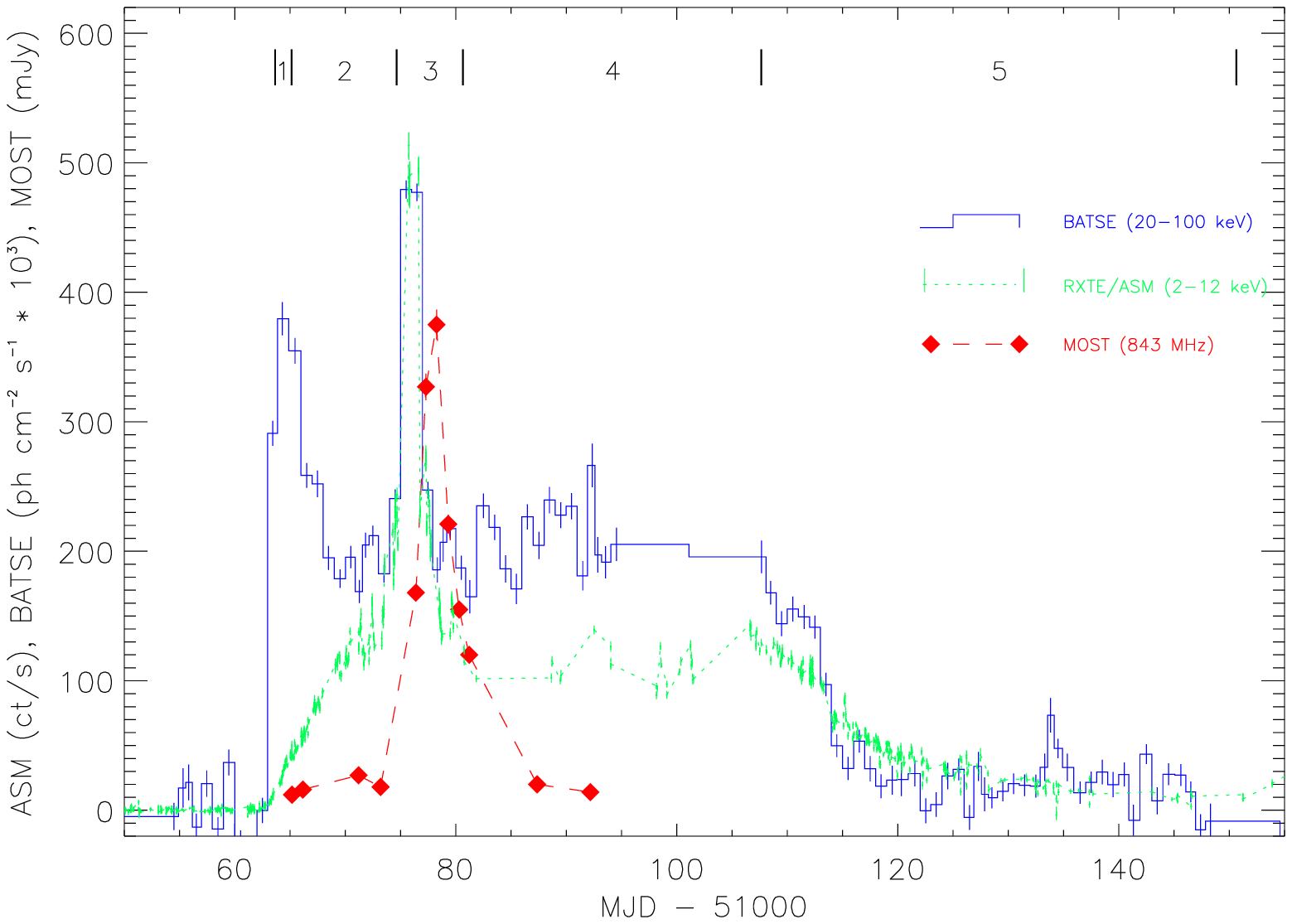,width=14.65cm}\\

\vspace{-0.6cm}
\hspace{0.4cm} \epsfig{figure=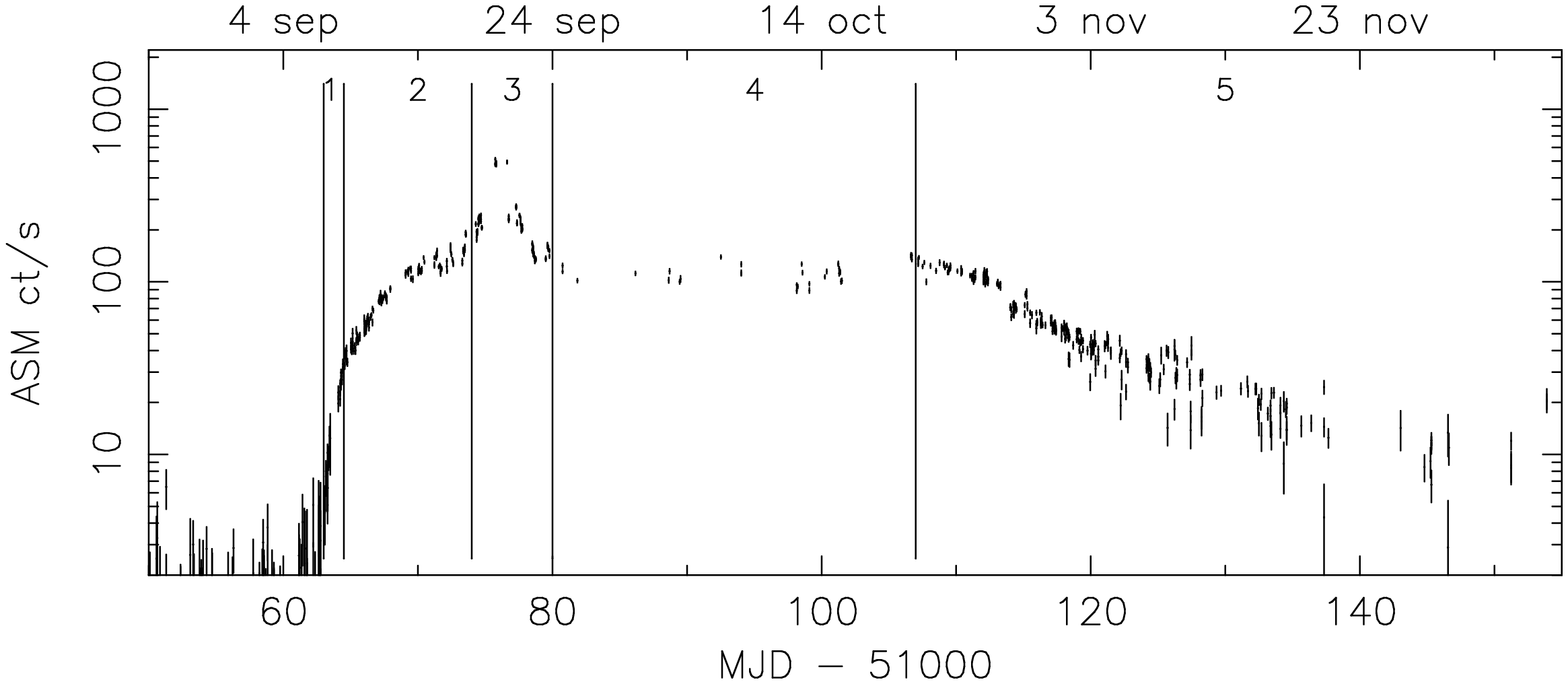,height= 13.0cm, angle=270}\\
\caption{Top panel: Optical/IR/UV lightcurves for the 1998 outburst. 
Open circles: data from Sanchez-Fernandez et al. (1999); 
filled circles: data from Jain et al. (1999). The upper lightcurve 
is in the standard $I$ band for Jain et al. (1999)'s dataset, and 
in Gunn $i$ for Sanchez-Fernandez et al. (1999)'s. The other two lightcurves 
are in standard $V$ and $U$ for both datasets. A possible 
sharp flare at MJD 51076 
was detected in the $V$ band by Sanchez-Fernandez et al. (1999), but was not 
seen by Jain et al. (1999).
Central panel: X-ray and radio lightcurves of \xte1550. 
Bottom panel: the phases of the 1998 outburst sequence 
shown on a logarithmic plot of the {\it RXTE}/ASM lightcurve. \label{fig1}}
\end{figure}

\begin{figure}
\epsfig{figure=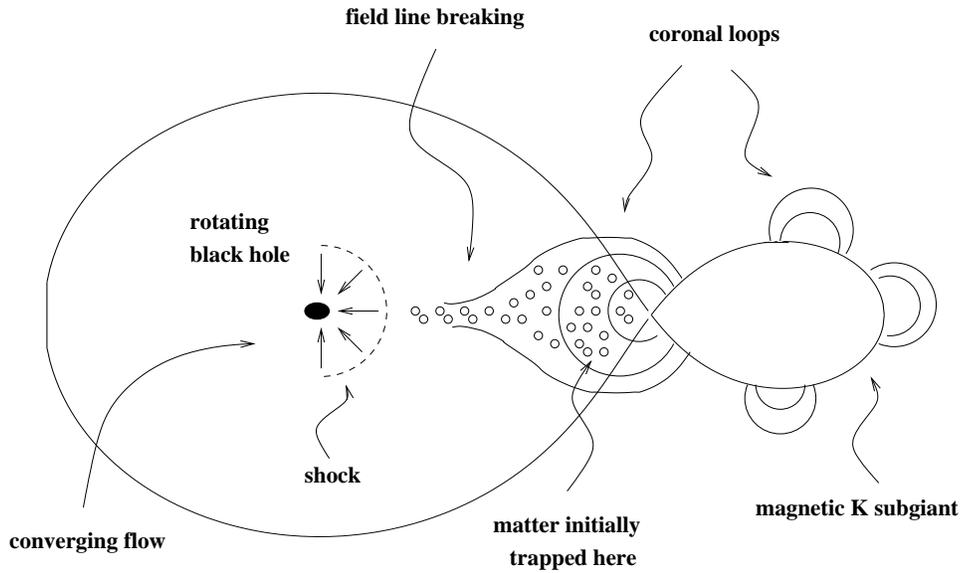,width=8.5cm,angle=270} 
\caption{A schematic illustration of the magnetic-star model 
      for \xte1550 at the onset of the 1998 outburst.  
   Matter is trapped inside the Roche lobe of the black hole, 
   near the inner Lagrangian point, 
      by the magnetic field of the K subgiant. This provides the initial 
low angular momentum component of the accretion flow. 
}
\end{figure} 


\begin{figure} 
\epsfig{figure=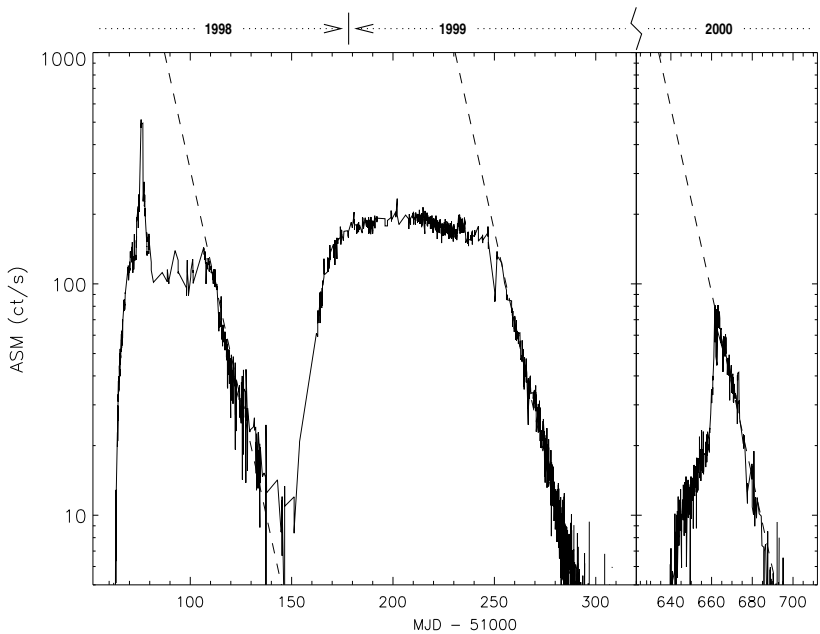,width=8.5cm,angle=270} 
\caption{The {\it RXTE}/ASM lightcurves 
     for the three outbursts of \xte1550 between 1998 and 2000. 
   The dashed lines correspond to an exponential decay 
     with an e-folding timescale of 11~days. }
\end{figure}

\begin{figure} 
\epsfig{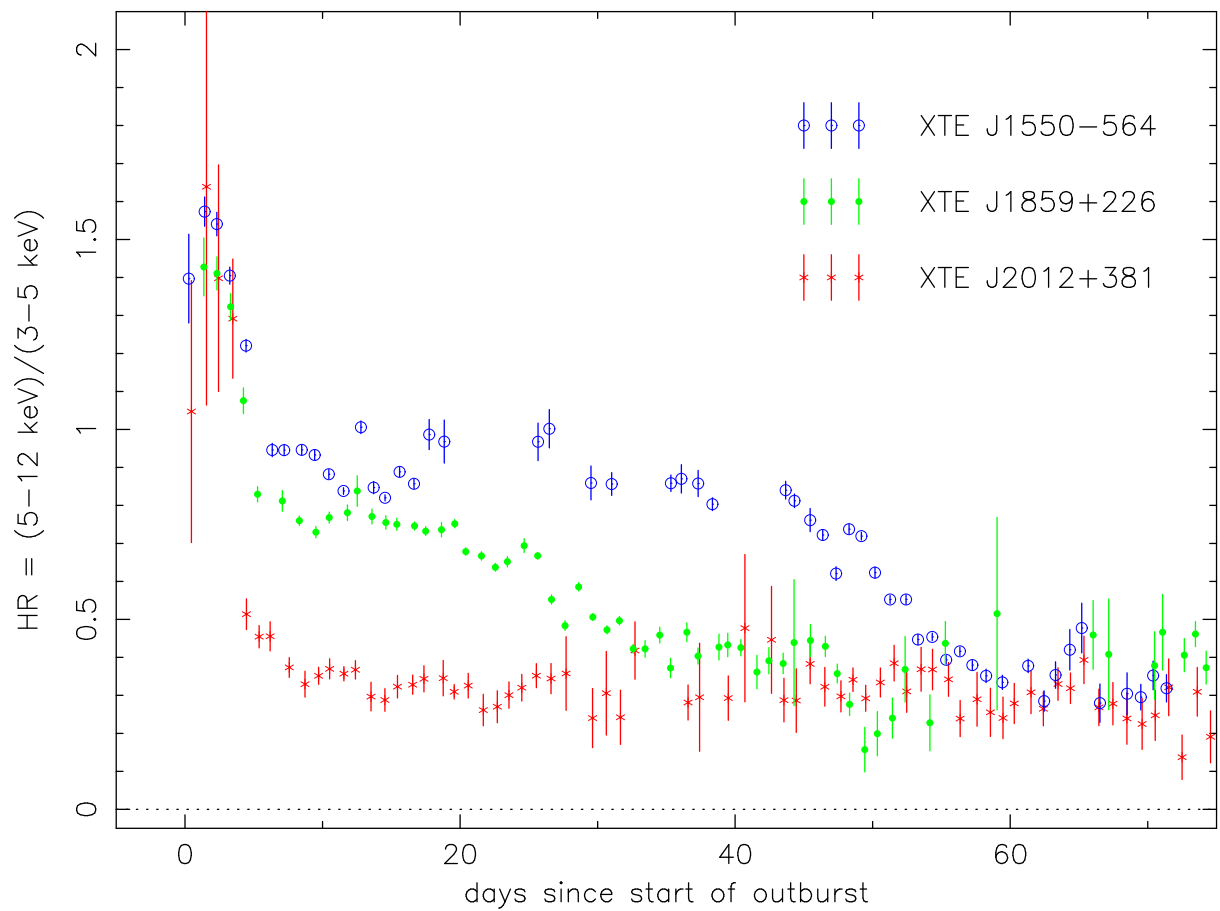} 
\caption{Evolution of 
the {\it RXTE}/ASM hardness ratios (counts in the 5--12 keV band / 
counts in the 3--5 keV band) during the outbursts of the BHCs \xte1550 
(open circles), 
XTE J1859$+$226 (filled circles) and XTE J2012$+$381 (asterisks). 
For \xte1550, time $=$ MJD $-$ 50063; 
for XTE J1859$+$226, time $=$ MJD $-$ 51459, and for XTE J2012$+$381, time $=$ 
MJD $-$ 50956. All three systems show an initial hard spike.}
\end{figure}

\end{document}